\documentclass[a4paper,fleqn]{cas-sc}

\usepackage[numbers]{natbib}

\def\tsc#1{\csdef{#1}{\textsc{\lowercase{#1}}\xspace}}
\tsc{WGM}
\tsc{QE}

\begin{document}
\let\WriteBookmarks\relax
\def\floatpagepagefraction{1}
\def\textpagefraction{.001}

\shorttitle{Exploring the expansion of the universe using the Gr\"uneisen parameter}

\shortauthors{Lucas Squillante, Gabriel O. Gomes, Isys F. Mello, Guilherme Nogueira, Antonio C. Seridonio, Roberto E. Lagos-Monaco, Mariano de Souza}

\title [mode = title]{Exploring the expansion of the universe using the Gr\"uneisen parameter}

\author[1]{Lucas Squillante}
\author[2]{ Gabriel O. Gomes}
\author[1]{ Isys F. Mello}
\author[1]{ Guilherme Nogueira}
\author[3]{ Antonio C. Seridonio}
\author[1]{ Roberto E. Lagos-Monaco}
\author[1]{ Mariano de Souza}
\fnmark[*]
\nonumnote{* Corresponding author email: mariano.souza@unesp.br}

\affiliation[1]{Sao Paulo State University (Unesp), IGCE - Physics Department, Rio Claro - SP, Brazil}
\affiliation[2]{University of Sao Paulo, Department of Astronomy, Sao Paulo, Brazil}
\affiliation[3]{Sao Paulo State University (Unesp), Department of Physics and Chemistry, Ilha Solteira - SP, Brazil}

\begin{abstract}
For a perfect fluid, pressure $p$ and energy density $\rho$ are related via the equation of state (EOS) $\omega = p/\rho$, where $\omega$ is the EOS parameter, being its interpretation usually constrained to a numerical value for each universe era. Here, based on the Mie-Gr\"uneisen EOS, we show that $\omega$ is recognized as the effective Gr\"uneisen parameter $\Gamma_{eff}$, whose singular contribution, the so-called Gr\"uneisen ratio $\Gamma$, quantifies the barocaloric effect. Our analysis suggests that the negative $p$ associated with dark-energy implies a metastable state and that in the dark-energy-dominated era $\omega$ is time-dependent, which reinforces recent proposals of a time-dependent cosmological constant. Furthermore, we demonstrate that $\Gamma_{eff}$ is embodied in the energy-momentum stress tensor in the Einstein field equations, enabling us to analyse, in the frame of an imperfect fluid picture, anisotropic effects of the universe expansion. We propose that upon going from decelerated- to accelerated-expansion, a phase transition-like behavior can be inferred. Yet, our analysis in terms of entropy, $\Gamma$, and a by us adapted version of Avramov/Casalini's model to Cosmology unveil hidden aspects related to the expansion of the universe. Our findings pave the way to interpret cosmological phenomena in connection with concepts of condensed matter Physics via $\Gamma_{eff}$.
\end{abstract}




\maketitle

\section{Introduction}\label{sec1}

\begin{figure}[!h]
\centering
\includegraphics[clip,width=\columnwidth]{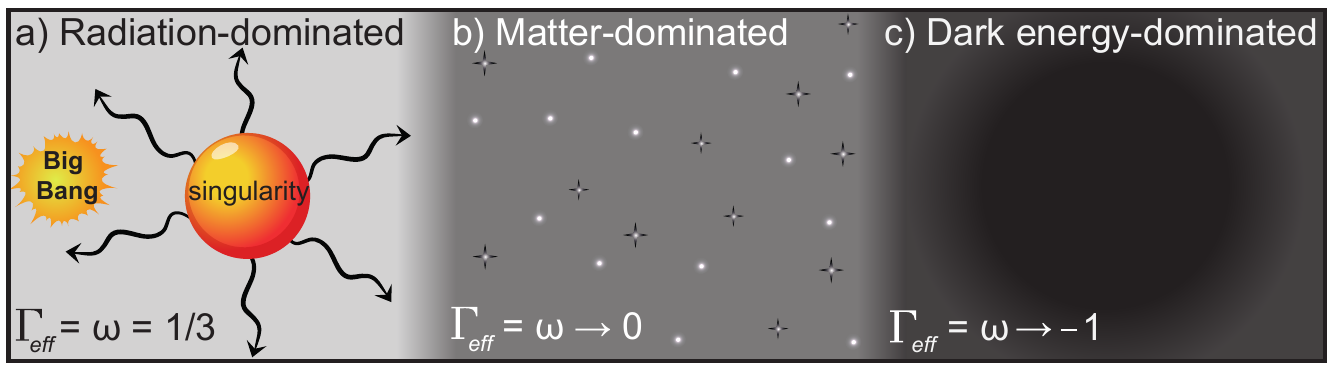}
\caption{\footnotesize Schematic representation of the various eras of the universe. In panel a), the radiation-dominated era ($\Gamma_{eff} =\omega = 1/3$), right after the Big Bang, is depicted. In b), the matter-dominated era is depicted ($\Gamma_{eff} = \omega \rightarrow 0$). In c), the dark-energy-dominated era is sketched, $\Gamma_{eff} = \omega \rightarrow -1$. The colour gradient represents the proposed role of the dark-energy in the accelerated expansion of the universe. Upon going from a) to c), the sign-change of $\Gamma_{eff}$ is reminiscent of a phase transition-like behaviour within a phases coexistence region close to a critical point, as it occurs in condensed matter Physics \cite{Barto}.}
\label{Fig-1}
\end{figure}

The observational evidence supporting the hypothesis that the universe is currently undergoing an accelerated expansion process is astonishing, see Fig.\,\ref{Fig-1} a-c) \cite{Kamionkowski2019}. The physical mechanism governing the dynamics of such accelerated expansion, as well as its physical origin remain under debate \cite{Kamionkowski2019}. Type Ia Supernova apparent magnitude measurements as a function of the redshift indicate accelerated cosmological expansion \cite{refs}. The obtaining of experimentally reliable cosmological parameters is of crucial importance to probe the validity of the various proposed theoretical models aiming to correctly predict the observed behavior of galaxy motions and apparent magnitude measurements. In the last few years, the Planck collaboration \cite{Planck2020} has provided precise measurements regarding cosmological parameters. Among the proposed models \cite{Bonometto2006,Wu2007,Grande2007,Neupane2008} to describe the expansion of the universe, the so-called $\Lambda$-Cold Dark Matter (CDM) model plays an important role, being CDM the only type of dark matter considered in such a model \cite{CDM}, which assumes the existence of a cosmological constant $\Lambda$. It has been discussed that $\Lambda$ governs the accelerated expansion process that counterbalances the gravitational attraction caused by ordinary matter \cite{Weinberg1972}. Essentially, the three contributions involved in the dynamics of the late-time universe expansion are: \emph{i}) CDM; \emph{ii}) dark energy (DE), which is implicitly taken into account by considering the existence of $\Lambda$; and \emph{iii}) ordinary matter.
The current universe is assumed to be dominated by $\Lambda$ with a time-independent energy density $\rho$ associated with DE \cite{Peebles1984}. Although the $\Lambda$-$\textrm{CDM}$ model has been extensively explored throughout the last decades, alternative models using the concept of a Dark Fluid (DF) have been proposed with the goal of unifying DE and dark matter (DM) \cite{Chavanis}. In this context, a number of EOS have been proposed. For instance, in Ref.\,\cite{Kamenshchik2001}, the authors proposed a model for a DF based on the so-called Chaplygin EOS. In this approach, the early stages of the universe are modeled by a pressureless fluid, i.e., a scenario in which DM governs the universe dynamics. The late stages of the cosmological expansion process are, following discussions in \cite{Kamenshchik2001}, governed by a fluid of constant $\rho$, being hence responsible for the current accelerated expansion of the universe. However, for the timespan between these two limiting cases, the predictions of the DF model based on the Chaplygin gas are in odd with experimental observations \cite{Zhu2004}, since the current available experimental data regarding the dimensionless coordinate distances and the X-ray gas mass fractions do not follow the EOS for the Chaplygin gas within a 99\% confidence level \cite{Zhu2004}. Furthermore, sophisticated DF models have been employed to properly reconcile experimental data with the theoretical description of the cosmological expansion \cite{Boshkayev2019}, such as the generalized and modified Chaplygin gas models \cite{Soares2006,Santos2007}. In short, there is a discrepancy between the models employed to describe cosmological expansion and the mechanisms governing the expansion throughout different eras. Here, we explore the Thermodynamics of an evolving universe with focus on the so-called effective Gr\"uneisen parameter \cite{Barto,Souza2015,EJP2016,SR2019,Mottiscool}, hereafter $\Gamma_{eff}$. We demonstrate that the so-called EOS parameter $\omega$ is recognized as $\Gamma_{eff}$ via Mie-Gr\"uneisen EOS \cite{Shalom1986}, which enables us to infer that $\Gamma_{eff}$ is embodied in the energy-momentum stress tensor $T_{\mu\nu}$ in Einstein field equations. We discuss the behaviour of $\Gamma_{eff}$ throughout the various eras employing the EOS of a perfect fluid and the modified Chaplygin gas. In our analysis we consider a flat universe scenario, i.e, $k=0$ in the Friedmann–Lema\^itre–Robertson–Walker (FLRW) metric of the Einstein field equations, except for the discussion regarding a phase-transition-like behaviour from decelerated- to accelerated-expansion in the frame of a closed spherical model for the universe.

\section{Thermodynamics of an expanding universe under the light of a perfect fluid approach}\label{sec3}

There is a long-standing debate in the literature about the applicability of equilibrium Thermodynamics to describe the expansion of the universe. While some authors claim that such treatment is erroneous, arguing that its continuous expansion is a non-equilibrium phenomenon \cite{Lavenda2015}, others argue that the time-scale associated with the cosmic expansion is much higher than the one associated with experimental observations, so that a thermodynamic equilibrium scenario can be inferred \cite{Astumian2006}. Under this perspective, we discuss in the following the Thermodynamics associated with the perfect fluid picture using $\Gamma_{eff}$. A general parametrization of the kind $p = p(\rho)$, where $p$ is pressure and $\rho$ the energy density, is usually employed to describe a homogeneous and isotropic universe, as well as to analyse the EOS for distinct periods of the universe inflation. For a perfect fluid, $p$ and $\rho$ are linked via $\omega$, the so-called EOS parameter, by the simple form $p = \omega \rho$ \cite{Weinberg1972}. For the radiation, matter, and DE dominated eras, $\omega$ = 1/3, 0, and $-$1, respectively \cite{Weinberg1972}. It has been discussed that the late stage of the universe is composed predominantly by vacuum energy ($\omega \rightarrow -1$), i.e., DE, where $\rho \propto a(t)^{-3(1+ \omega)}$, and $-1\le \omega <0$ \cite{Huterer2008} for the various eras, being $a$ the scale factor. Considering Friedmann equations \cite{Friedman}, the relation between $\rho$, $p$, and the specific entropy $\sigma$ reads \cite{Weinberg1972}:
\begin{equation}
nTd\sigma=d\rho - \frac{p + \rho}{n}dn,
\label{sigma}
\end{equation}
where $n$ is the number of particles per volume and $T$ temperature. From Eq.\,\ref{sigma}, the temporal evolution of the temperature for a perfect fluid is given by \cite{Calvao2002} (see supplementary material):
\begin{equation}
\frac{\dot{T}}{T} = \left(\frac{\partial p}{\partial \rho}\right)_n \frac{\dot{n}}{n}.
\label{temperatureevolution}
\end{equation}
Interestingly enough, the EOS for a perfect fluid is reminiscent of the well-known Mie-Gr\"uneisen EOS, namely $p_a = \Gamma_{eff}E_a/V$, where $p_a$ and $E_a$ refer, respectively, to the contributions of the atomic vibrations to $p$ and energy, which is broadly employed in condensed matter Physics for solids and fluids \cite{Gruneisen1926,Shalom1986,Chernyshev2011}. Evidently, in the case of fluids, $\Gamma_{eff}$ links $p$ and the kinetic/thermal energy of the particles. For solids, $\Gamma_{eff}$ weights the ratio between pressure and energy when $V$ is changed upon varying a tuning parameter, e.g., $T$ or stress. Unambiguously, based on the above discussions, it becomes evident that $\omega$ plays the role of $\Gamma_{eff}$ in the EOS of a perfect fluid in the description of the universe expansion. Thus, under the light of a perfect fluid-based approach, $\Gamma_{eff}$ dictates how $p$ and $\rho$ are affected in terms of the temporal evolution of the scale factor for each era, which in turn governs the expansion of the universe. One of the key-results of the present work refers to the perspective that $\omega$ is not simply an EOS parameter, i.e., it is not merely a proportionality factor linking $p$ and $\rho$, as it will become clear in the following. Integrating Eq.\,\ref{temperatureevolution} we achieve the well-known expression (see supplementary material):
\begin{equation}
TV^{\Gamma_{eff}} = constant,
\label{scalingrelation}
\end{equation}
which relates $T$ and $V$ with $\Gamma_{eff}$ during an adiabatic expansion of the universe in the frame of a perfect fluid analysis \cite{Dalgarno2005,Weinberg1970}. Thus, it is clear that its temperature is continuously reduced as it expands, which in turn can be interpreted as a barocaloric effect \cite{Mariano2021}. The latter is quantified upon employing the singular contribution to $\Gamma_{eff}$, the so-called Gr\"uneisen ratio $\Gamma$, namely \cite{EJP2016,Zhu2003}:
\begin{equation}
\Gamma = \frac{1}{V_m T}\left(\frac{\partial T}{\partial p}\right)_{S},
\label{GR}
\end{equation}
where $V_m$ and $S$ refer to the molar volume and the entropy, respectively. In condensed matter Physics, the first step for the realization of a barocaloric effect is the application of pressure, followed by its adiabatic removal, so that cooling is achieved. In the case of the universe, the singularity before the Big Bang could be considered as the first step for the realization of a barocaloric effect. Considering the inherent uncertainty in the estimation of $T$ and $p$ of the expanding universe at each era \cite{Melendres2021}, a quantification of such a (baro)caloric effect in terms of $\Gamma$ is precluded. In the radiation- and matter-dominated eras, $\Gamma_{eff} > 0$ and, as a consequence, the temperature of the universe decreases to compensate its expansion, so that the adiabatic character is obeyed. However, counterintuitively, during the DE-dominated era, $T$ increases during the adiabatic expansion, i.e., $\Gamma_{eff} < 0$, cf.\,Eq.\,\ref{scalingrelation}, which is the principle of an inverse barocaloric effect \cite{Majumdar2011}. This holds true only for the portion associated with the proposed DE, which, in turn, can dictate the temperature evolution of the universe in its late stages \cite{Komatsu2020}. At this point, we recall the relation between $\Gamma_{eff}$ and $\Gamma$ \cite{Zhu2003}:
\begin{equation}
\Gamma_{eff} = \Gamma\frac{c_p}{c_v}\frac{V}{\kappa_T}.
\end{equation}\newline
Given that $c_p/c_v = \kappa_S/\kappa_T$ \cite{earth}, where $\kappa_S$ and $\kappa_T$ are, respectively, the adiabatic and isothermal compressibilities, $\Gamma_{eff}$ reads:
\begin{equation}
\Gamma_{eff} = \frac{\Gamma \kappa_S V}{{\kappa_T}^2} \Rightarrow \Gamma = \frac{\Gamma_{eff}{\kappa_T}^2}{\kappa_S V}.
\label{gammaeffandgamma}
\end{equation}
Employing the canonical definition of $\kappa_S$ and $\kappa_T$ \cite{earth}, the fact that $\Gamma_{eff} = \omega$, and $(\partial V/\partial p)_S = -V/\omega\rho$ from the perfect fluid EOS, we achieve a relation between $\Gamma$ and $\omega$, namely $\Gamma_{eff}$, for the matter- and radiation-dominated eras in the frame of a perfect fluid:
\begin{equation}
\Gamma = \frac{\omega}{pV}.
\label{gammaomega}
\end{equation}
Hence, since $\omega = \Gamma_{eff} = 1/3$ and 0 $\Rightarrow$ $\Gamma = 1/3pV$ and 0 for the radiation- and matter-dominated eras, respectively. Given that the temperature evolution of the universe can be interpreted as a barocaloric effect, which in turn is quantified by $\Gamma$, our analysis enables us to infer that the adiabatic temperature variation of the universe was continuously diminished upon approaching the matter-dominated era. For the DE-dominated era, the peculiarity of $p < 0$ is crucial and must be taken into account, so that for $\omega = \Gamma_{eff} = -1$ $\Rightarrow$ $\Gamma = -(pV)^{-1}$, which is consistent with an inverse barocaloric effect associated with the DE-dominated era, as well as with our analysis for the adiabatic expansion considering $\omega = \Gamma_{eff} = -1$, cf.\,Eq.\,\ref{scalingrelation}.

Now, we make an analysis in terms of the volume- and temperature-dependence of $S$ for the various eras. At this point, we recall that we are dealing with an adiabatic expansion, so that $dS = 0$. Using $\Gamma_{eff} = v\alpha_p/\kappa_Tc_v$ \cite{EJP2016} and the thermodynamic relation $\alpha_p = \kappa_T(\partial S/\partial V)_T$, it is straightforward to obtain:\newline
\begin{equation}
V\left(\frac{\partial S}{\partial V}\right)_T = \Gamma_{eff} c_v.
\end{equation}
Since $\Gamma_{eff} = \omega$ and employing $c_v = T(\partial S/\partial T)_V$, we have:
\begin{equation}
V\left(\frac{\partial S}{\partial V}\right)_T = \omega T \left(\frac{\partial S}{\partial T}\right)_V.
\end{equation}
Thus, for the various eras:
\begin{equation}
V\left(\frac{\partial S}{\partial V}\right)_T = \frac{T}{3} \left(\frac{\partial S}{\partial T}\right)_V, \hspace{1cm}\textmd{(radiation-dominated era)}
\label{rad}
\end{equation}\vspace{-0.3cm}
\begin{equation}
\hspace{0.35cm}\left(\frac{\partial S}{\partial V}\right)_T \rightarrow 0, \hspace{2.35cm}\textmd{(matter-dominated era)}
\label{mat}
\end{equation}
\begin{equation}
\hspace{-0.1cm}V\left(-\frac{\partial S}{\partial V}\right)_T = T \left(\frac{\partial S}{\partial T}\right)_V. \hspace{1.05cm}\textmd{(DE-dominated era)}
\label{de}
\end{equation}\newline
A simple analysis of Eq.\,\ref{rad} enables us to infer that in the radiation-dominated era, the isothermal entropy variation due to the universe expansion was more pronounced than the one associated with its isochoric temperature-dependent variation by a factor 3. On the verge of the regimes change from decelerated- to accelerated-expansion, $S$ remains approximately constant, cf.\,Eq.\,\ref{mat}, which is in perfect agreement with our analysis of Eq.\,\ref{gammaomega} for the matter-dominated era. Considering that $(\partial S/\partial V)_T = p/T$ \cite{Lavenda2015}, Eq.\,\ref{de} only holds true upon assuming negative pressures. Hence, the assumption of a negative pressure associated with DE is recovered upon analysing Eq.\,\ref{de}. Yet, it is straightforward to infer that there is a balance between the volume and the temperature dependence of $S$. At this point, it is worth exploring the consequences of the consideration of a negative $p$ associated with DE. As a matter of fact, the concept of negative $p$ and its link with a metastable state was nicely discussed by Landau \cite{Landau1980}. Thus, based on Ref.\,\cite{Landau1980}, $(\partial S/\partial V)_T < 0$ implies in a metastable state, where a spontaneous contraction of the system, in this case the volume portion associated with DE, takes place and, as a consequence, $S$ is increased due to the possible formation of cavities \cite{Lavenda2015,Landau1980}. Following discussions in Ref.\,\cite{Melendres2021}, the contraction of the volume portion associated with DE can explain the accelerated expansion in the DE-dominated era, since the so-called Hubble flow is increased due to such a contraction. Although there are some approaches considering the possibility of associating the concept of negative temperatures to the DE-dominated era \cite{phantom1,phantom2,phantom3}, in some cases called phantom regime, this is questionable since, as nicely discussed in Ref.\,\cite{ramsey}, negative temperatures require thermodynamic equilibrium as it is found, for instance, in the canonical definition of temperature based on the magnetic energy of a paramagnetic system \cite{unveiling}. At this point, given our recognition of $\omega$ as $\Gamma_{eff}$, we discuss the various contributions to $\omega$. In condensed matter Physics, the various contributions to $\Gamma_{eff}$, namely, magnetic, phononic, and electronic, must be taken into account, so that \cite{EJP2016}:
\begin{equation}
\Gamma_{eff} = \frac{\sum_i \Gamma_{eff, i}\,c_{v_i}}{\sum_i c_{v_i}},
\end{equation}
where $c_v$ is the specific heat at constant volume. The index $i$ represents the $i^{th}$ contribution to $\Gamma_{eff}$ and $c_v$. Going beyond the simple picture of a perfect fluid, i.e., in the frame of a multicomponent universe \cite{Weinberg1972}, we propose that $\omega$, namely $\Gamma_{eff}$, can also be expressed as the sum of its various distinct contributions, namely:
\begin{equation}
\omega = \frac{\sum_i \omega_i\,\rho_i}{\sum_i \rho_i},
\label{dominance}
\end{equation}\newline
where $\omega_i$ and $\rho_i$ refer to the contributions coming from the radiation-, matter-, and DE-dominated eras. Note that, following Eq.\,\ref{dominance}, the ultimate value of $\omega$ depends on the dominance of a particular component of $\rho$ for each era. Just for the sake of completeness, it is worth mentioning that considering an universe composed by an ideal gas, $\omega = \Gamma_{eff} = 2/3$ \cite{EJP2016,arxivpereira}. Now, we make use of the approach of a closed spherical universe \cite{silk,Melendres2021} aiming to analyse its speed of expansion in connection with $\Gamma_{eff}$. Assuming that the universe is an isolated system and that its expansion is adiabatic, we recall the first-law of Thermodynamics:
\begin{equation}
dQ = dU + dW,
\end{equation}
where $dQ$, $dU$, and $dW$ are the infinitesimal heat, internal energy, and work increments. Since $dQ = TdS$ and $dW = +pdV$, i.e., the universe is performing work on the hypothetical spherical wall of the ``container'' at the expense of its internal energy in order to expand, we have \cite{Melendres2021}:
\begin{equation}
TdS = dU + pdV,
\end{equation}
which for an adiabatic process leads to $dU = -pdV$ \cite{Melendres2021}. In a Newtonian-like approach, considering that the Hubble parameter $H = \dot{a}/a$ \cite{Knox2022} is the normalized rate of expansion, $H$ can be analogously determined in terms of the normalized rate of the temporal evolution of the universe's volume, namely \cite{Melendres2021}:
\begin{equation}
H = \frac{\dot{V}}{V} = \frac{\frac{dV}{dt}}{\frac{4}{3}\pi r^3},
\label{hubble}
\end{equation}
where $r$ is the radius of the hypothetical spherical universe discussed at this point. Equation\,\ref{hubble} can thus be rewritten as \cite{Melendres2021}:
\begin{equation}
H = \frac{3}{4\pi r^3} \frac{dV}{dU}\frac{dU}{dt},
\end{equation}
so that \cite{Melendres2021,Busca2013}:
\begin{equation}
\frac{dU}{dt} = \frac{4}{3}\pi r^3 H \frac{dU}{dV}.
\label{dedt}
\end{equation}\newline
Following discussions in Ref.\,\cite{Busca2013}, during the radiation-dominated era $dU/dt < 0$, i.e.,  the universe is expanding in a decelerated way. Upon reaching the matter-dominated era, the speed of expansion reaches a minimum, i.e., $dU/dt \rightarrow 0$, and then, as it enters in the DE-dominated era, $dU/dt > 0$ indicating that the universe is expanding in an accelerated way. Since $dU = -pdV$ and the volume variation is positive due to the universe expansion, negative $p$ implies that the internal energy of the universe is increasing in the DE-dominated era \cite{Melendres2021}. Hence, a connection between $\Gamma_{eff}$, the speed of expansion of the universe, and the temporal evolution of $U$ can be made. As previously discussed, for an adiabatic expansion $dU/dV = -p$ and since for the perfect fluid $p = \omega\rho = \Gamma_{eff}\rho$ $\Rightarrow$ $dU/dV = -\Gamma_{eff}\rho$. Hence, Eq.\,\ref{dedt} can be rewritten as:
\begin{equation}
\Gamma_{eff} = \frac{3}{4\pi r^3 H \rho}\left(-\frac{dU}{dt}\right).
\label{lambdacdm}
\end{equation}
Assuming that $r$, $H$, and $\rho$ are always positive, the sign of $\Gamma_{eff}$ is governed solely by $-dU/dt$. Thus, in the regimes $dU/dt < 0$ $\Rightarrow$ $\Gamma_{eff} > 0$; $dU/dt \rightarrow 0$ $\Rightarrow$ $\Gamma_{eff} \rightarrow 0$, and for $dU/dt > 0$ $\Rightarrow$ $\Gamma_{eff} < 0$. Such a simple analysis enables us to recover the behaviour of $\Gamma_{eff}$ throughout the various eras, cf.\,Fig.\,\ref{Fig-1}. The sign-change of $\Gamma_{eff}$ upon going from decelerated- to an accelerated-expansion emulates the case of a phase transition close to a critical endpoint in condensed matter Physics \cite{Mottiscool}. Based on Noether's theorem \cite{Noether}, the change in the time dependence of $U$ upon going from the decelerated to accelerated periods can be interpreted as a symmetry breaking, which is analogous to the case of a canonical phase transition \cite{Landau1980}. Also, it is worth mentioning that it is under debate in the literature whether $\omega$ has indeed converged to $-$1 in the DE-dominated era or if it would change over time as the universe keeps expanding \cite{Planck2020}. Under the light of the here-proposed recognition of $\omega$ as $\Gamma_{eff}$, it is tempting to infer that eventually there would be a time-dependence of $\omega$, namely $\Gamma_{eff}$, since the thermodynamic quantities embedded in $\Gamma_{eff}$, such as, $\kappa_T$ and $c_v$, will vary as a consequence of the continuous temperature change and expansion of the universe. Hence, since $\omega$, namely, $\Gamma_{eff}$, in the perfect fluid approach is linked with $\Lambda$ via $p = \omega\rho_{\Lambda} = -\rho_{\Lambda} = -c^2\Lambda/8\pi G$ \cite{Weinberg1972}, where $c$ is the speed of light, $\rho_{\Lambda}$ the vacuum energy density, and $G$ the gravitational constant, the result of a time-dependent $\omega$ extracted from our analysis implies in a $\Lambda(t)$ or $G(t)$ (or both), as proposed in some recent works \cite{pimentel,tutusaus,shukla}. To clarify such an issue, accurate experimental observations of $\rho_{\Lambda}$ over time are highly desirable \cite{frieman}. In the following, we employ the modified Chaplygin gas EOS as a working horse to demonstrate that the identification of $\omega = \Gamma_{eff}$ is not restricted to the perfect fluid EOS.

\section{The modified Chaplygin gas}\label{sec4}

The EOS of the modified Chaplygin gas reads \cite{Santos2007}:
\begin{equation}
p = B\rho - \frac{A}{\rho^{\alpha}},
\label{MCG}
\end{equation}
where $A$, $B$, and $\alpha$ are non-universal constants, being $0 < \alpha \leq 1$ \cite{Santos2007}. In the limit of the early stage of the universe, $\rho \rightarrow \infty$ and thus, Eq.\,\ref{MCG} becomes simply $p \simeq B\rho$, which recovers the EOS for a perfect fluid with $B$ playing the role of $\omega$. As the universe expands, $\rho$ is continuously reduced and thus both terms of the right side of Eq.\,\ref{MCG} contribute to $p$. Using the relation $(dU/dV)_S = -p$ and the corresponding expression for the internal energy, please refer to Eq.\,5 of Ref.\,\cite{Santos2007}, $\rho$ and $p$ reads:
\begin{equation}
\rho = \left(\frac{A}{B+1}\right)^{1/(\alpha+1)}\left[1+\left(\frac{\delta}{V}\right)^{R^*}\right]^{1/(\alpha+1)},
\end{equation}
\begin{equation}
p = -\left(\frac{A}{B+1}\right)^{1/(\alpha+1)}\frac{(B+1)}{[1+(\delta/V)^{R^*}]^{\alpha/(\alpha+1)}}\left\{1 - \frac{B}{B+1}\left[1+\left(\frac{\delta}{V}\right)^{R^*}\right]\right\},
\label{pressuremcg}
\end{equation}
where $\delta = [b^*(B+1)/A]^{1/R^*}$, $b^*$ is an integration constant, and $R^* = (B+1)(\alpha+1)$ \cite{Santos2007}. A characteristic volume is defined as $V^* = \delta B^{1/R^*}$ \cite{Santos2007}, where $V < V^*$ accounts for positive pressures and $V > V^*$ negative ones \cite{Lavenda2015}. Essentially, $V < V^*$ mimics the radiation-dominated era while $V = V^*$ the matter-dominated era, and $V > V^*$ the DE dominated one. For the early stages of the universe, i.e., $V \ll \delta$, $\rho$ and $p$ reads \cite{Santos2007}:
\begin{equation}
\rho \approx \left(\frac{A}{B+1}\right)^{1/(\alpha+1)}\left(\frac{\delta}{V}\right)^{(B+1)},
\end{equation}
\begin{equation}
p \approx B\left(\frac{A}{B+1}\right)^{1/(\alpha+1)}\left(\frac{\delta}{V}\right)^{(B+1)}.
\end{equation}\newline
Rewriting $p$ in terms of $\rho$ and based on the fact that $\omega =$ $\Gamma_{eff}$, we have then for $V \ll \delta$, $\Gamma_{eff} \approx B$, which restores the result for the perfect fluid when $B = 1/3$. Thus, for the early stage of the universe, the modified Chaplygin gas recovers nicely $\Gamma_{eff}$ for the perfect fluid. Also, for $V = V^*$, $p = 0$ and thus $\Gamma_{eff} = 0$ as well, which is in perfect agreement with the result for the matter-dominated era employing the EOS of a perfect fluid. For late times, i.e., $V \gg \delta$, so that $\rho$ and $p$ reads \cite{Santos2007}:
\begin{equation}
\rho \approx \left(\frac{A}{B+1}\right)^{1/(\alpha+1)}\left[1+\frac{1}{\alpha+1}\left(\frac{\delta}{V}\right)^{R^*}\right],
\end{equation}
\begin{equation}
p \approx -\left(\frac{A}{B+1}\right)^{1/(\alpha+1)}.
\end{equation}
In such a regime, $\rho$ behaves as a mixing of two fluids \cite{Santos2007}. Employing $\rho(p$), $\Gamma_{eff}$ can be computed:
\begin{equation}
\Gamma_{eff} = - \frac{(\alpha+1)V^{R^*}}{[(\alpha+1)V^{R^*} + \delta^{R^*}]}.
\label{gammaeffmcg}
\end{equation}
Since we are dealing with the case of $V \gg \delta$, $\Gamma_{eff} \rightarrow -1$, which is in line with the result for the DE dominated era employing the EOS for a perfect fluid. Note that $\delta$ introduces the mixture character between the EOS of a perfect fluid and the Chaplygin gas. Interestingly, Eq.\,\ref{gammaeffmcg} has some reminiscence with the results obtained by some of us for the van der Waals EOS, namely $\Gamma_{eff} = (2/3)V/(V-Nb)$ \cite{EJP2016}, where $N$ is the number of particles and $b$ the volume associated with them. Note that for $V \gg Nb$ $\Rightarrow$ $\Gamma_{eff} \rightarrow 2/3$, which recovers $\Gamma_{eff}$ for the ideal gas \cite{EJP2016}. In the present case, $\delta^{R^*}$plays a role analogous to $Nb$, since when $V \gg \delta^{R^*}$ $\Rightarrow$ $\Gamma_{eff} \rightarrow -1$, so that the result for the DE-dominated era in the frame of the perfect fluid picture is nicely restored. Indeed, the analogy between the ideal gas and perfect fluid holds true since both have the same $T_{\mu\nu}$ \cite{Anderson1967}. Our analysis of the universe expansion in the frame of the modified Chaplygin gas considering a finite $\delta$ in terms of $\Gamma_{eff}$ is in line with the proposal of a ``cosmological'' mixture composed by DE and DM. In other words, the mixture of the two contributions to $p$ embedded in the modified Chaplygin gas EOS, cf.\,Eq.\,\ref{MCG}, is captured in the analysis of $\Gamma_{eff}$ (Eq.\,\ref{gammaeffmcg}) in terms of the parameter $\delta^{R^*}$. Furthermore, the consideration of a finite value of $\delta^{R^*}$ is crucial since $\omega$, namely $\Gamma_{eff}$, will never be exactly $-$1, as expected for the DE-dominated era in the frame of a perfect fluid. Making an analogy with the van der Waals gas, cf.\,previous discussion, and, considering that the universe is expanding adiabatically, it is natural to expect that the value of $\delta^{R^*}$ will become smaller and smaller than $V$, making, eventually, $\Gamma_{eff} \rightarrow -1$.

\section{Gr\"uneisen parameter and the Einstein field equations: anisotropic expansion effects}

It is well-known that the celebrated Einstein field equations relate the curvature of spacetime in terms of a generalized distribution of energy and momentum. This set of equations is given by \cite{Einstein1916}:
\begin{equation}
R_{\mu\nu} - \frac{1}{2}g_{\mu\nu}R + \Lambda g_{\mu\nu} = 8\pi G T_{\mu\nu},
\label{EFE}
\end{equation}
where $R_{\mu\nu}$ is the Ricci curvature tensor, $g_{\mu\nu}$ the metric tensor, $R$ the curvature scalar. For a perfect fluid, the term $T_{\mu\nu}$ is related with $p$ and $\rho$ by $T_{\mu\nu} = (p+\rho)u_{\mu}u_{\nu}+pg_{\mu\nu}$ \cite{Weinberg1972,Dalarsson2015,Shang2005}, where $u$ is the four-velocity vector field. Thus, using the EOS of a perfect fluid it is straightforward to write $T_{\mu\nu}$ as a function of $\omega$ as $T_{\mu\nu} = (1+\omega)\rho u_{\mu}u_{\nu}+\rho\omega g_{\mu\nu}$ \cite{Weinberg1972}. Since $\omega = \Gamma_{eff}$, we rewrite unprecedentedly the Einstein field equations in terms of $\Gamma_{eff}$ employing the EOS of a perfect fluid, namely:
\begin{equation}
R_{\mu\nu} - \frac{1}{2}g_{\mu\nu}R + \Lambda g_{\mu\nu} = 8\pi G[(1+\Gamma_{eff})\rho u_{\mu}u_{\nu}+\rho\Gamma_{eff} g_{\mu\nu}].
\label{gammaeinstein}
\end{equation}
Hence, it becomes clear that the Einstein field equations naturally embody $\Gamma_{eff}$ through $T_{\mu\nu}$. To the best of our knowledge, up to date, there is no discussion in the literature about the Einstein field equations in connection with the Gr\"uneisen parameter. Note that the link between $T_{\mu\nu}$ and $\Gamma_{eff}$ is given by the fact that $T_{\mu\nu}$ embodies pressure components, shear stresses and the energy density associated with the gravitational field. Such terms/components dictate the relation between $p$ and $\rho$ via the EOS. In other words, the consideration of particular terms in $T_{\mu\nu}$ establishes the form of the EOS. Our analysis regarding the connection between $\Gamma_{eff}$ and $T_{\mu\nu}$ is of particular interest considering an imperfect fluid, which represents one of the possible scenarios to explain current observational models that deviate from the $\Lambda$-CDM model by considering additional anisotropic energy fluxes and the stress tensor components \cite{Algoner2019}. In this picture, the effective stress energy-momentum tensor is $T_{\mu\nu} = [T_{(per)\mu\nu} + T_{(imp)\mu\nu}]$ \cite{Algoner2019}, where the indexes $(per)$ and $(imp)$ refer to the contributions from perfect and imperfect fluids, respectively. The analysis in terms of an imperfect fluid model is in line with recent experimental observations regarding an anisotropic expansion of the universe, which in turn can be seen as a direct consequence of the formation of galaxy clusters with different sizes in the early stages \cite{Lovisari2020}. For the sake of completeness, it is worth mentioning that the anisotropic expansion has been probed via measurements of the redshift space distortions \cite{Huff2022}. Also, for the DE-dominated era, it was reported in the literature that quantum vacuum is very inhomogeneous and anisotropic, being $\nabla a$ and $\nabla^2a$ expressive \cite{unruh}. At this point, we employ similar arguments than those reported in Ref.\,\cite{Mottiscool} by some of us regarding the metal-insulating coexistence region of the Mott transition aiming to describe the anisotropic expansion of the universe. It turns out that the dynamics of particles in inner galaxy clusters depends on particular aspects of a specific considered galaxy cluster \cite{wolfendale}. In Ref.\,\cite{Mottiscool}, using Avramov/Casalini's model as a working horse, we assumed different relaxation times for metallic and insulating bubbles embedded in the coexistence region of the Mott metal-to-insulator transition. As reported in Refs.\,\cite{Barto,prl2010}, upon crossing the first-order transition line within the coexistence region of the Mott transition, anisotropic effects in the thermal expansion are observed. An analogous situation can be considered for the universe assuming different periods of rotation for distinct galaxy clusters. In this analogy, the galaxies clusters play the role of the metallic/insulating bubbles. Upon going from radiation-, passing by matter-, to DE-dominated era, $\Gamma_{eff}$ changes sign, a behavior analogous to that observed in the pressure-induced Mott metal-to-insulator transition close to a second-order critical endpoint, cf.\,Ref.\,\cite{Mottiscool}. Thus, based on the fact that the scale factor varies over time, an anisotropic expansion of the universe can be inferred as well. As a matter of fact, the proposal of the presence of critical points related to the universe expansion was discussed in Ref.\,\cite{Melendres2021}. It is clear that anisotropic expansion effects can be described in terms of stress tensor components, which in turn can be associated with the elastic Gr\"uneisen parameter recently proposed by some of us \cite{Mariano2021}, namely $\Gamma_{ec} = 1/T(\partial T/\partial \sigma_{ij})_S$, where $\sigma_{ij}$ represents the stress tensor components. Note that $\Gamma_{ec}$ equals $\Gamma$ (Eq.\,\ref{GR}) when $p$ is rewritten in terms of $\sigma_{ij}$. Hence, since $\Gamma_{eff}$ is related to $\Gamma$, cf.\,Eq.\,\ref{gammaeffandgamma}, anisotropic expansion effects can be explored in terms of $\Gamma_{ec}$ embedded in $\Gamma_{eff}$, which in turn is incorporated in $T_{\mu\nu}$ in Einstein field equations. Considering that DE drives accelerated expansion without ``reins'', one of the ultimate fates of the universe would be the proposed ``big rip'', a topic still under debate in the literature, see, e.g., Refs.\,\cite{Weinberg2003,katie,skibba}. Such a scenario can be treated in terms of $\Gamma_{ec}$, which incorporates all stress tensor components and is embodied in $T_{\mu\nu}$ in the Einstein field equations. Hence, we propose an alternative path to describe the possible scenario of a ``big rip'' of the universe under the light of Condensed Matter Physics.
Upon considering additional density components in $T_{{\mu}{\nu}}$, the Einstein field equations should be solved in order to explore the effect of such additional density components in the expansion of the universe. In our case, we have employed the simple picture of the perfect fluid to recognize $\Gamma_{eff}$ in Einstein field equations, which shall be explored in more details in future works. Now, we make an analysis of the various eras using the approach proposed by Casalini using Avramov's model for glasses \cite{casalini,paluch,avramov} adapted by us to Cosmology. We start recalling the expression for the relaxation time $\tau$ \cite{Mottiscool,casalini2}:
\begin{equation}
\tau = \tau_0\exp{\left(\frac{C}{TV^{\Gamma_{eff}}}\right)},
\label{tau}
\end{equation}
where $C$ is a non-universal constant and $\tau_0$ a reference relaxation time. Note that $\Gamma_{eff}$, namely $\omega$, dictates the behavior of $\tau$. Here, we propose a characteristic expansion time to each cluster of galaxies, which we assume, in turn, to be related to the frequency $f$ distribution of their rotating masses \cite{plionis}. We recall that a cluster of galaxies can be formed by a few thousands of galaxies \cite{plionis,hughes}. In other words, our assumption is, following Avramov's model for glasses, that there is a distribution of $f$ inherent to the existence of distinct rotating galaxies clusters masses. Hence, the ratio $\tau/\tau_0 \propto f_0/f$, where $f_0$ is a characteristic frequency of such distribution of cluster masses. It is evident that $f$ is linked to a corresponding period of rotation, so that $\tau$ emulates the behaviour of such a rotation period. Thus, under such assumptions, employing Eq.\,\ref{tau} considering $\Gamma_{eff}$ for the various eras, we have:
\begin{equation}
\frac{\tau}{\tau_0} = \exp{\left(\frac{C}{TV^{1/3}}\right)},\hspace{0.6cm} \textmd{(radiation-dominated era)}
\label{tau1}
\end{equation}
\begin{equation}
\frac{\tau}{\tau_0} = \exp{\left(\frac{C}{T}\right)},\hspace{1.3cm}\textmd{(matter-dominated era)}
\label{tau2}
\end{equation}
\begin{equation}
\hspace{-0.1cm}\frac{\tau}{\tau_0} = \exp{\left(\frac{CV}{T}\right)}.\hspace{1.1cm} \textmd{(DE-dominated era)}
\label{tau3}
\end{equation}\newline
In terms of rotation period, our analysis reveals that, in the radiation-dominated era, the period of the rotating galaxies clusters was reduced as the universe expands, cf.\,Eq.\,\ref{tau1}, while in the matter-dominated era such a period was $V$-independent, cf.\,Eq.\,\ref{tau2}. In the DE-dominated era, in turn, the period of the rotating galaxies clusters is exponentially increased as the universe expands, cf.\,Eq.\,\ref{tau3}. Our analysis suggests an expressive slowing-down of the dynamics of the rotating galaxies clusters in the DE-dominated era, being, astonishingly, a similar situation found on the verge of the Mott metal-insulator transition \cite{Mottiscool}. Hence, it is clear that $\omega$, here recognized as $\Gamma_{eff}$, plays a key role in the dynamics of rotating galaxies clusters throughout the eras.

\section{Conclusion}\label{sec13}

We have identified the EOS parameter $\omega$ as $\Gamma_{eff}$ via Mie-Gr\"uneisen EOS and analysed it for the perfect and imperfect fluids, as well as for the modified Chaplygin gas. Our analysis suggests that in the DE-dominated era $\omega$ is time-dependent, which implies that $\Lambda$ varies over time. We have proposed that the barocaloric effect associated with the adiabatic universe expansion, as well as the anomalous behaviour of DE regarding an inverse barocaloric effect, can be quantified by the Gr\"uneisen ratio. We have shown that $\Gamma_{eff}$ changes sign upon going from decelerated- to accelerated-expansion, indicating a phase transition-like behaviour. Our findings suggest that the temperature of the universe was diminished adiabatically upon approaching the matter-dominated era, while for the DE-dominated era our analysis points to an inverse barocaloric effect associated with DE. We propose that negative pressures associated with DE necessarily imply in a metastable state and that negative temperatures usually associated with the phantom phase may be erroneous due to the lack of a thermal equilibrium associated with the universe continuous expansion. Employing an adapted version of Avramov/Casalini's model to Cosmology, the dynamics of rotating galaxies clusters was analysed in terms of $\Gamma_{eff}$ for the various eras. We have shown that $\Gamma_{eff}$ is naturally embedded in $T_{\mu\nu}$ in the Einstein field equations. In this regard, our analysis provides an alternative path to describe anisotropic effects associated with the expansion of the universe under the light of $\Gamma_{eff}$, which is connected with the possibility of a ``big rip''. We suggest that our analysis in terms of $\Gamma_{eff}$ can be extended to other metrics, e.g., the Kerr metric, corresponding to a rotating uncharged axially-symmetric black hole \cite{Kerr1963}. It is challenging to revisit the Thermodynamics of black holes and related topics of Cosmology using $\Gamma_{eff}$. Our findings open a new avenue in the field of Cosmology to interpret the expansion of the universe, aiming to contribute to the current necessity of new approaches and interpretations to account for the limitations of the current standard cosmological scenario, cf.\,highlighted in Ref.\,\cite{review}.

\section*{Declaration of competing interests}

The authors declare that they have no known competing financial interests or personal relationships that could have appeared to influence the work reported in this paper.

\section*{Data availability}

Data will be made available on request.

\section*{Acknowledgements}
MdeS acknowledges financial support from the S\~ao Paulo Research Foundation – Fapesp (Grants number 2011/22050-4, 2017/07845-7, and 2019/24696-0), National Council of Technological and Scientific Development – CNPq (Grants number \,302887/2020-2). ACS acknowledges National Council of Technological and Scientific Development – (Grants No. 308695/2021-6). This work was partially granted by Coordena\c c\~ao de Aperfeiçoamento de Pessoal de N\'ivel Superior - Brazil (Capes) - Finance Code 001 (Ph.D. fellowship of L.S. and I.F.M.).







\end{document}